\documentclass{PoS}

\title{Recent STAR Measurements to Constrain the Polarized Gluon Distribution Function of the Proton}

\ShortTitle{Gluon polarization measurements at STAR}

\author{\speaker{Amilkar Quintero}\thanks{For the STAR collaboration.}\\
        Temple University\\
        E-mail: \email{amilkar.quintero@temple.edu}}

\abstract{The STAR experiment has been studying the spin structure of the proton, using the unique high-energy polarized proton collider, the Relativistic Heavy Ion Collider (RHIC). The kinematic coverage at STAR allows accessing gluons because quark-gluon and gluon-gluon scatterings dominate particle production at low and medium transverse momenta. The polarized gluon distribution function can be constrained by measuring the longitudinal double-spin asymmetry ($A_{LL}$) of jet production and neutral pions. Global QCD analyses of polarized parton distribution functions, which include the 2009 $A_{LL}$ STAR results for inclusive jet production at $\sqrt{s}=200$ GeV, provide evidence of a non-zero gluon polarization in the measured range of partonic momentum fraction $x > 0.05$. We present the status of the latest measurements of $A_{LL}$ at STAR, for inclusive jet and dijet production at $\sqrt{s}=510$ GeV and 200 GeV collected in 2013 and 2015 respectively, both at mid-rapidity ($|\eta|<0.9$). The large data sample taken during these years will improve the precision of our knowledge about the proton spin structure while the increased center of mass energy allows probing the polarized gluon distribution function at smaller partonic momentum fraction. Furthermore, we present measurements of $A_{LL}$ for neutral pions at forward rapidity ($2.65<\eta<3.9$) collected during 2012 and 2013 that also allow reaching lower partonic momentum fraction. We compare these measurements with the latest global analyses.}

\FullConference{XXVI International Workshop on Deep-Inelastic Scattering and Related Subjects (DIS2018)\\
		16-20 April 2018\\
		Kobe, Japan}

\begin{document}

\section{Introduction}
	One of the major goals of the Relativistic Heavy Ion Collider (RHIC) spin program is to perform high precision measurements of the polarized gluon distribution function ($\Delta g(x)$) of the proton. These RHIC measurements are motivated by the results of polarized Deep Inelastic Scattering (DIS) in the late 1980's, that showed quarks contribute partially to the spin of the proton. Recent measurements have shown that the quarks account for approximately 25\% of the spin of the proton (see \cite{review} and references therein), corroborating that quarks alone are not able to explain the magnitude ($1/2$) of the spin of the proton, so gluons might be responsible of some of the remaining amount. 
	
	RHIC, located at Brookhaven National Laboratory, is the only polarized proton-proton collider in the world with center of mass energies up to 510 GeV. For the spin program, a polarized hydrogen source is used to provide polarized protons. These protons are pre-accelerated in several stages before entering the RHIC rings. The polarization of the proton bunches is constantly monitored during each stage. Siberian Snakes are used to counteract the depolarization of the protons. Spin rotators are installed before and after each interaction point to provide longitudinal polarization, and also they can provide independent spin polarization choice for each experiment at RHIC. The polarized proton bunches are collided with different spin patterns to avoid any related systematic effects \cite{polarized}. 

	The Solenoidal Tracker At RHIC (STAR) is one of the largest experiments at RHIC. The main tracking device is a Time Projection Chamber (TPC) with full azimuthal coverage and $|\eta| \le 1.3$. Electromagnetic calorimeters ($-1 \le \eta \le 2$) are used to trigger on high momentum particles and measure the neutral component of jets. The Forward Meson Spectrometer (FMS) is a lead-glass EM calorimeter to detect $\pi^{0}$ at $2.5 \le \eta \le 4.2$. The Vertex Position Detector (VPD, a set of scintillators) and the Zero Degree Calorimeter (ZDC) are used to measure the relative luminosity; both detectors are located at both sides of the interaction point at very forward pseudo-rapidity \cite{STAR}.

\section{Gluon polarization measurements}

	 For most RHIC kinematics, $gg$ and $qg$ scattering dominate, making RHIC accessible to probe polarized gluons at high energies. Gluon polarization can be measured using longitudinally polarized double spin asymmetry ($A_{LL}$) of jets and neutral pions. The STAR experiment is well equipped to perform jet reconstruction. Jet measurements are insensitive to the hadronization processes, so jets are a good probe to study spin physics and constrain the gluon polarization distrbution function. Despite excellent $A_{LL}$ results in previous years, there is still the need to increase the precision in the sampled momentum fraction ($x$) region and extend $x$ to lower values.

\subsection{Inclusive Jets}

	Previous STAR inclusive jet $A_{LL}$ results, at the center of mass energy $\sqrt{s}=200$ GeV \cite{pibero}, provide the first evidence of positive gluon polarization. The 2009 results were systematically above earlier global fits by the DSSV and NNPDF collaborations, suggesting a positive gluon polarization value. After inclusion of these STAR results, the gluon polarization value obtained was $\Delta G=0.19 \pm 0.06$ ($0.05<x$) \cite{DSSV14} and $\Delta G=0.23_{-0.07}^{+0.06}$ ($0.05<x<0.5$) \cite{NNPDF11}, from updated global analyses independently performed by the aforementioned groups. The uncertainties for this measurement, at low momentum fraction, are still very large, so the sensitivity needs to be extended to lower $x$ to further constrain global fits.
	
\begin{figure}
\begin{center}
\includegraphics[scale=0.51]{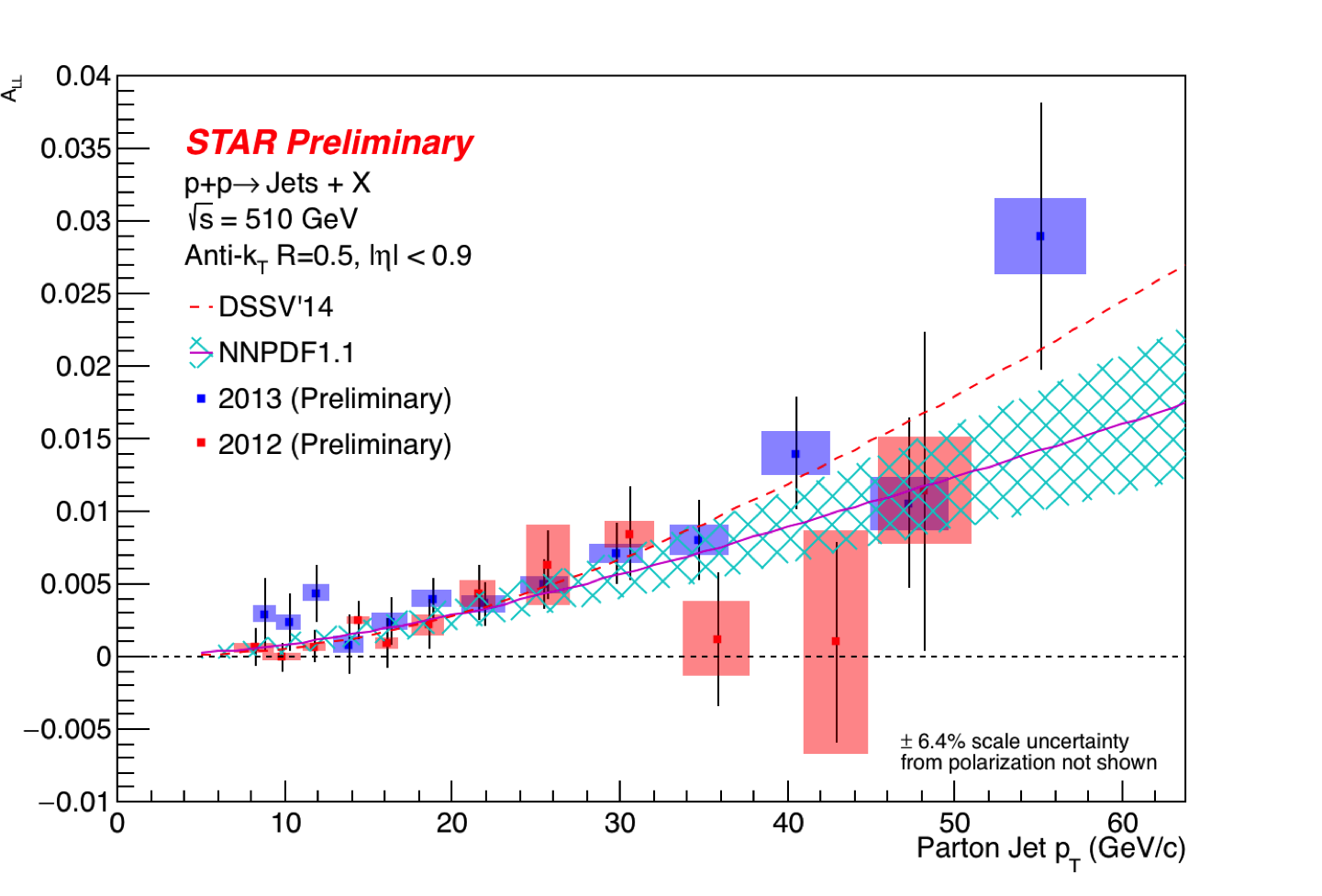}
\end{center}
\caption{\label{fig:one}Preliminary results for 2012 \cite{zilong} and 2013 inclusive jets $A_{LL}$ as a function of parton jet $p_{T}$, compared to DSSV14 \cite{DSSV14} and NNPDFpol1.1 \cite{NNPDF11} predictions. Vertical lines represent the statistical uncertainties, while blue and red boxes show the systematic uncertainties.}
\end{figure}

	In 2012 and 2013 STAR collected data at the center of mass energy $\sqrt{s}=510$ GeV. The increased center of mass energy allows probing lower momentum fraction values down to $x > 0.02$. Figure \ref{fig:one} shows the STAR preliminary results for inclusive jets $A_{LL}$ as a function of the parton jet transverse momentum from 2012 data \cite{zilong} and the latest preliminary result from 2013 data.
	
	Jets were reconstructed using the anti-kt jet finding algorithm \cite{antikt} with $R=0.5$. Simulations embedded in data are used to quantify the detector response and estimate systematic uncertainties. The figure of merit relevant for double spin asymmetry in 2013 is almost three times greater than the previous year, due to the increased luminosity provided by the collider in that year. The STAR collaboration installed the Heavy Flavor Tracker (HFT) \cite{hft} in the middle of the 2013 run; the extra material affects the reconstruction of the jets, therefore the preliminary results for the 2013 run only include data prior to the installation of the HFT, which represent approximate $60\%$ of the total statistics available. The embedded simulation sample for the 2012 run was used for systematic uncertainty estimations. The inclusion of the STAR run 2009 results to the newest global fits in 2014 provides better control of the systematics effects (e.g. trigger and reconstruction bias), allowing improvement of the related uncertainties.

\begin{figure}
\begin{center}
\includegraphics[scale=0.51]{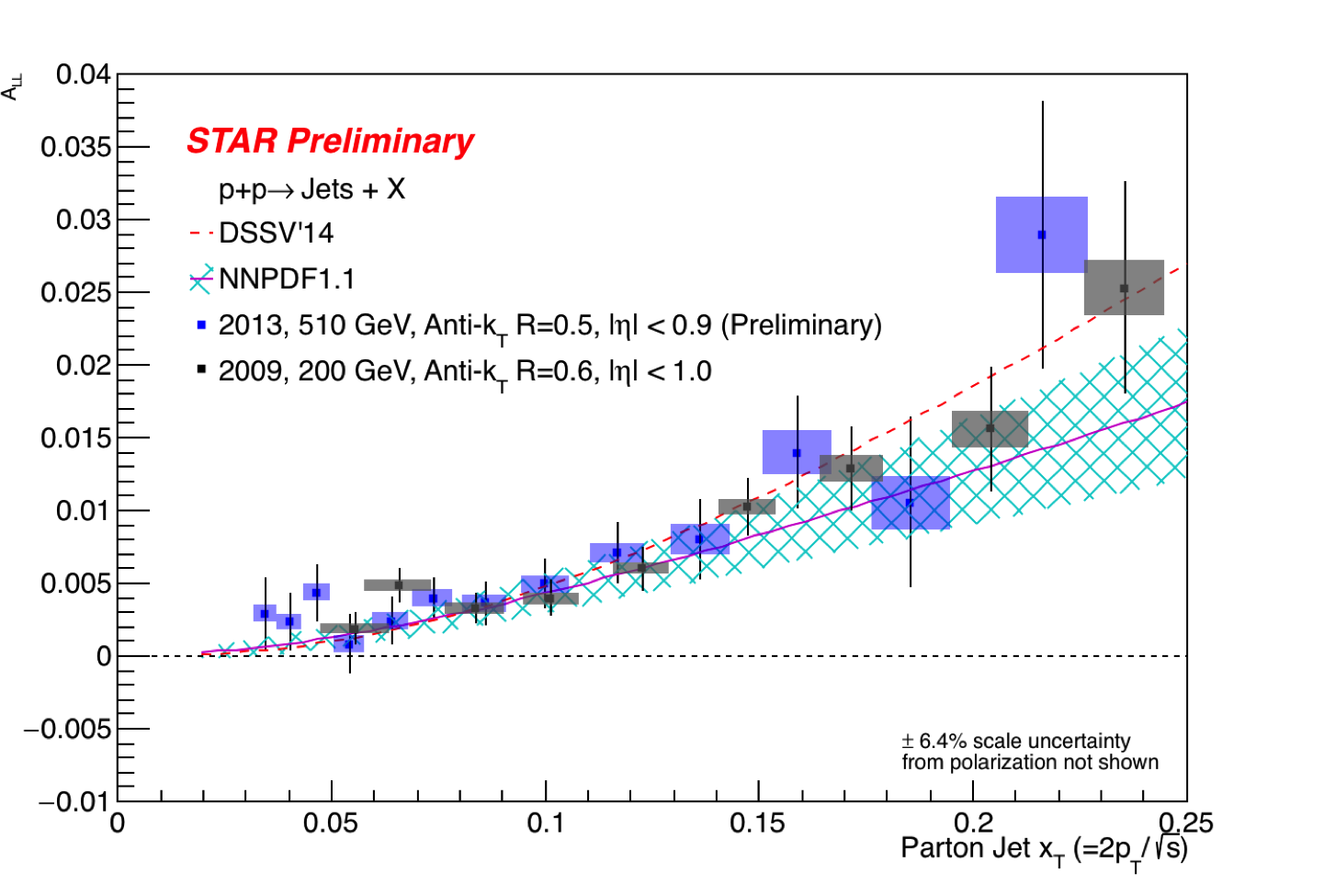}
\end{center}
\caption{\label{fig:two}Comparison between 2009 STAR result \cite{pibero} and 2013 preliminary result in the overlapping $x_{T}$ region, of inclusive jets $A_{LL}$, compared to DSSV14 \cite{DSSV14} and NNPDFpol1.1 \cite{NNPDF11} predictions. Vertical lines represent the statistical uncertainties, while blue and gray boxes show the systematic uncertainties.}
\end{figure}

	The $A_{LL}$ results for run 2009 (200 GeV), run 2012 and the newest run 2013 (510 GeV) show good agreement. Figure \ref{fig:two} shows the STAR 2009 inclusive jets $A_{LL}$ results \cite{pibero} and the 2013 preliminary results in the overlap $x_{T}$ region. The full data sample of run 2013 is already processed and simulation is being produced to finalize systematic uncertainty studies.

	In 2015, STAR collected a new set of data at the center of mass energy $\sqrt{s}=200$ GeV to increase the precision in the currently sampled region and consolidate the observation of a positive gluon polarization. These measurements will reduce uncertainties by a factor of approximately 1.6, compared to the 2009 measurement. The procedure to perform these $A_{LL}$ measurements at STAR is already well established, and the analysis for the 2015 data set is being finalized for a preliminary release.

\subsection{Di-jets}

\begin{figure}
\begin{center}
\includegraphics[scale=0.51]{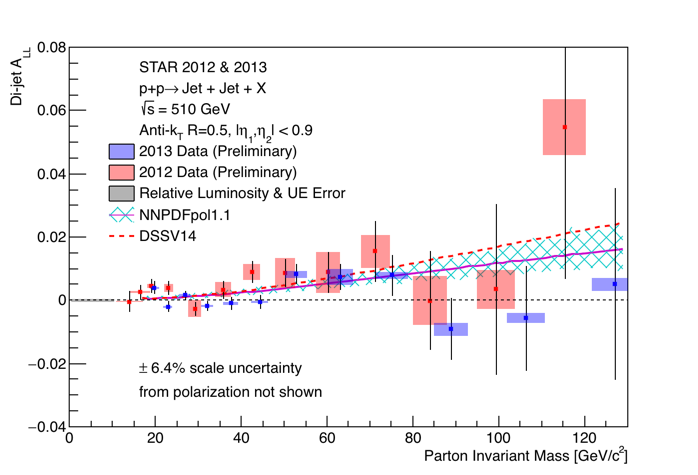}
\end{center}
\caption{\label{fig:three}Preliminary results for 2012 \cite{subarna} and 2013 runs of di-jet $A_{LL}$ vs. parton invariant mass compared to DSSV14 \cite{DSSV14} and NNPDFpol1.1 \cite{NNPDF11} predictions. Vertical lines represent the statistical uncertainties, while blue and red boxes show the systematic uncertainties.}
\end{figure}

\begin{figure}
\begin{center}
\includegraphics[scale=0.75]{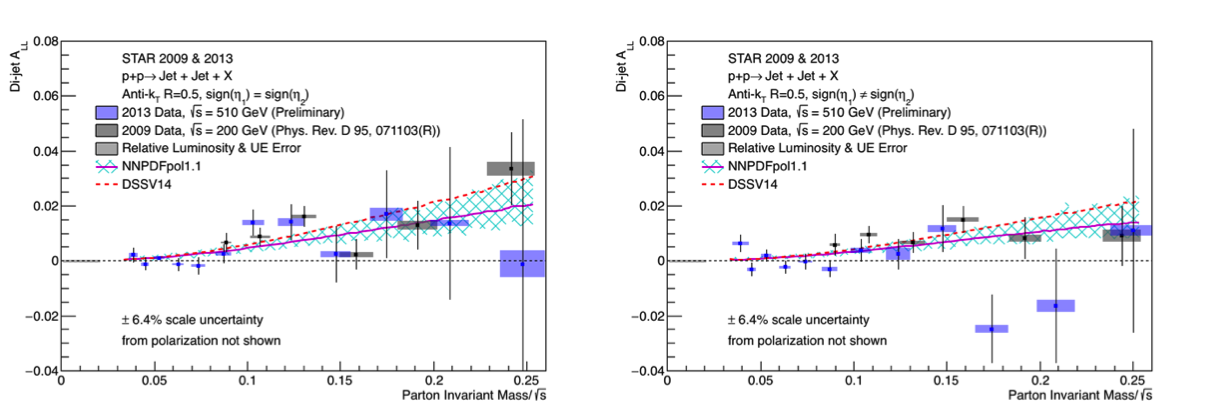}
\end{center}
\caption{\label{fig:four}Comparison between 2009 STAR result \cite{brian} and 2013 preliminary result, in the scaled parton invariant mass overlap region, of Di-jet $A_{LL}$ for same-sign (left) and opposite-sign (right) topological configurations measured; compared to DSSV14 \cite{DSSV14} and NNPDFpol1.1 \cite{NNPDF11} predictions. Vertical lines represent the statistical uncertainties, while blue and grey boxes show the systematic uncertainties.}
\end{figure}

	In 2009 STAR also performed di-jets measurements at the center of mass energy $\sqrt{s}=200$ GeV. Di-jets allow probing a narrower momentum fraction region than inclusive jets (see \cite{brian} figure 3). The 2009 STAR di-jet $A_{LL}$ results \cite{brian} are consistent with 2014 global fits. Additionally, the 2009 studies provided a cross section measurement of di-jet production, confirming that NLO pQCD calculations can correctly describe the data. Also in 2009, a forward di-jet $A_{LL}$ measurement at $-0.8< \eta < 1.8$ was done. This measurement allows to reach both lower $x$ and sample a narrower region. The STAR 2009 forward di-jet $A_{LL}$ results were recently submitted for publication \cite{ting}. 
	Similar to the inclusive jets, measurements of di-jet at a center of mass energy $\sqrt{s}=510$ GeV, were done. Figure \ref{fig:three} shows the STAR preliminary result for di-jet $A_{LL}$ as a function of the parton di-jet invariant mass from 2012 data \cite{subarna} and the latest preliminary result from 2013 data. Preliminary di-jet asymmetry results for 2009, 2012 and 2013 are in agreement with each other and with the 2014 global fits. Reduced statistical and systematic uncertainties from 2013 data compared to 2012 data were obtained, similarly for the inclusive jet studies. Preliminary results are in agreement with 2009 results \cite{brian} in the overlap region. Figure \ref{fig:four} shows the 2009 STAR results for di-jets $A_{LL}$ and the latest preliminary result in 2013, in the overlap region, presented in two topological configurations: when both jets have same-sign and opposite-sign pseudorapidities. Both inclusive jet and di-jet 2012 final results are in preparation for publication.

\subsection{Forward neutral pion measurements}
	The $A_{LL}$ measurements of forward neutral pions at the center of mass energy $\sqrt{s} = 510$ GeV allow reaching the lowest momentum fraction values at STAR. During the runs in 2012 and 2013, STAR operated the FMS for the spin program. The FMS consists of an array of lead-glass electromagnetic calorimeter cells, coupled to photomultiplier tubes. The FMS is divided in an inner region of smaller cells ($3.8\,cm \; x \; 3.8\,cm$), while the outer regions has larger cells   ($5.8\,cm \; x \; 5.8\,cm$). The STAR forward neutral pion $A_{LL}$ result was recently submitted for publication \cite{chris}.

\section{Summary}

	The STAR experiment has performed measurements of inclusive jets ($|\eta|<1$) and di-jets ($-0.8<\eta<1.8$) at $\sqrt{s}=200$ GeV and 510 GeV, and neutral pions ($2.65<\eta<3.9$) at $\sqrt{s}=510$ GeV. The STAR spin program provided, for the first time, evidence of positive gluon polarization. Results of $A_{LL}$ for inclusive jet, di-jets and $\pi^{0}$ are consistent with each other and in agreement with the global fits DSSV14 \cite{DSSV14} and NNPDFpol1.1 \cite{NNPDF11}. The 2013 run embedding studies are ongoing, so the path to final results is well established after completion of this Monte Carlo sample. The STAR collaboration took an additional $\sqrt{s}=200$ GeV polarized proton data set during 2015, to consolidate previous measurements in 2009. These high precision measurements motivate the natural step forward to the STAR forward upgrade program \cite{forward} and an Electron Ion Collider \cite{EIC}.

\end{document}